\documentclass{ifacconf}

\usepackage{natbib}        % required for bibliography\
\usepackage{graphicx}  
\usepackage{amsmath}
\usepackage{amssymb}
\usepackage{mathrsfs}
\usepackage{setspace}
\usepackage{epstopdf}
\usepackage{color}
\usepackage{booktabs}
\usepackage{multirow}
\usepackage{subfigure}
\usepackage{mathrsfs}
\usepackage{soul}
\usepackage{makecell}

\usepackage{algorithm}
\usepackage{algpseudocode}
\algnewcommand{\Input}[1]{\Statex\textbf{Input:} #1}
\algnewcommand{\Parameter}[1]{\Statex\textbf{Parameter:} #1}
\algnewcommand{\Output}[1]{\Statex\textbf{Output:} #1}

\newtheorem{definition}{Definition}

\newtheorem{lemma}{Lemma}

\begin{document}
\begin{frontmatter}
\title{{Economic data-enabled predictive control using machine learning}\thanksref{footnoteinfo}}

% \thanks[footnoteinfo]{This research is supported by\ Ministry of Education, Singapore, under its Academic Research Fund Tier 1 (RG63/22), and Nanyang Technological University, Singapore (Start-Up Grant).\\  $^{1}$ 
\thanks[footnoteinfo]{{This research is supported by\ PUB, Singapore’s National Water Agency under its RIE2025 Urban Solutions and Sustainability (USS) (Water) Centre of Excellence (CoE) Programme, awarded to Nanyang Environment \& Water Research Institute (NEWRI), Nanyang Technological University, Singapore (NTU), and Ministry of Education, Singapore, under its Academic Research Fund Tier 1 (RG63/22). Any opinions, findings and conclusions or recommendations expressed in this material are those of the author(s) and do not reflect the views of National Research Foundation, Singapore and PUB, Singapore's National Water Agency.}\\  $^{1}$ 
\hspace{0.5mm}Corresponding author: Xunyuan Yin. Tel: (+65) 6316 8746. Email: xunyuan.yin@ntu.edu.sg}

\author[First,Second,Third]{Mingxue Yan} 
\author[Third]{Xuewen Zhang} 
\author[Fourth]{Kaixiang Zhang}
\author[Fourth]{Zhaojian Li}
\author[First,Third]{Xunyuan Yin}$^{,1}$

\address[First]{Nanyang Environment and Water Research Institute, Nanyang Technological University, 1 CleanTech Loop, 637141, Singapore}  
% (e-mail: yanm0009@e.ntu.edu.sg; xunyuan.yin@ntu.edu.sg)
\address[Second]{Interdisciplinary Graduate Programme, Nanyang Technological University, 61 Nanyang Drive, 637460, Singapore}
% (e-mail: yanm0009@e.ntu.edu.sg)}
\address[Third]{School of Chemistry, Chemical Engineering and Biotechnology, \mbox{Nanyang Technological University, 62 Nanyang Drive, 637459, Singapore}}
% (e-mail: yanm0009@e.ntu.edu.sg; xuewen001@e.ntu.edu.sg; xunyuan.yin@ntu.edu.sg)}
\address[Fourth]{Department of Mechanical Engineering, Michigan State University, East Lansing, MI 48824, USA }
% (e-mail: zhangk64@msu.edu; lizhaojl@egr.msu.edu) }

\begin{abstract} 
In this paper, we propose a convex data-based economic predictive control method within the framework of data-enabled predictive control (DeePC). Specifically, we use a neural network to transform the system output into a new state space, where the nonlinear economic cost function of the underlying nonlinear system is approximated using a quadratic function expressed by the transformed output in the new state space. Both the neural network parameters and the coefficients of the quadratic function are learned from open-loop data of the system. 
Additionally, we reconstruct constrained output variables from the transformed output through learning an output reconstruction matrix; this way, the proposed economic DeePC can handle output constraints explicitly. The performance of the proposed method is evaluated via a case study in a simulated chemical process.
\end{abstract}


\end{frontmatter}
%===============================================================================

\section{Introduction}

Optimizing process operation performance, such as maximizing economic profits or minimizing economic costs, has been one of the primary objectives of optimal process control. One representative framework to address economic considerations is to rely on a two-layer hierarchical architecture~(\cite{marlin1997real}). The upper layer solves steady-state economic optimization. The optimal steady-state values serve as reference points for real-time control in the lower layer to track. Set-point tracking model predictive control (MPC) has been widely adopted to handle the real-time reference tracking tack in the lower layer~(\cite{limon2018nonlinear,rawlings2000tutorial}).
% ~(\cite{rawlings2000tutorial}). 
Meanwhile, the optimal process set-point can vary during process operations. 
% due to various factors, such as fluctuations in the cost of raw materials and energy prices. 
Frequent changes in set-points may compromise the control performance, as set-point tracking MPC requires time to have the process operation reach a new steady-state~(\cite{ellis2014tutorial}). 
Moreover, while steady-state operation is common in the industry, it may not deliver economically optimal results~\cite{ellis2014tutorial}. 
% These limitations have constrained the ability of set-point tracking MPC to optimize overall economic performance.

Economic model predictive control (EMPC)
% , which integrates real-time optimization with dynamic control in a unified framework,
offers a promising approach for effectively managing the economic performance of industrial systems and processes~(\cite{ellis2014tutorial, ellis2017economic}).
% ~(\cite{ellis2014tutorial}). 
In EMPC, a general cost function, which represents the economic cost or profit was used as the control objective function.
In~\cite{diehl2010lyapunov}, an EMPC method with point-wise terminal constraints was proposed.
In~\cite{heidarinejad2012economic}, Lyapunov-based constraints were incorporated in the EMPC framework to guarantee closed-loop stability. 
In~\cite{amrit2011economic}, the stability of the EMPC framework was guaranteed with terminal cost. 
In~\cite{grune2014asymptotic}, the stability of EMPC without terminal conditions was studied. EMPC has also been applied across various industries. 
In~\cite{hovgaard2013nonconvex}, a nonconvex EMPC design was developed for a commercial refrigeration system and demonstrated a sophisticated response to real-time variations in electricity price.  
% Substantial cost savings on the order of 30\% 
% were achieved, and sophisticated response to real-time variations in electricity price was observed.
While EMPC addresses the limitations of set-point tracking MPC, it can be hindered by the lack of accurate first-principles models. Exceptions can be found in~\cite{albalawi2023koopman, han2024efficient}. In~\cite{albalawi2023koopman}, Koopman modeling was carried out to construct a linear model based on which an EMPC scheme was developed. In~\cite{han2024efficient}, a learning-based input-output Koopman model was established. A convex EMPC problem was formulated by training a quadratic economic cost function.
% A neural network was learned to facilitate the construction of a quadratic function that approximates the economic cost in a lifted state space, and a convex EMPC design was developed for nonlinear systems.
Meanwhile, most of the existing EMPC methods require full-state measurements for online control implementation, which has hindered their widespread adoption across industries. This limitation arises because obtaining real-time measurements of all key process state variables can often be impractical~\cite{yin2019subsystem,yin2018estimation}. 
% In real applications, obtaining online measurements of all key variables through hardware sensors may be infeasible~\cite{yin2019subsystem}. 

% An emerging framework, which is referred to as the Koopman-based EMPC, has attracted academic attention due to its ability to cope with the absence of high-fidelity models~\cite{han2024efficient,albalawi2023koopman}. According to the Koopman theory, a surrogate linear model can potentially be found for any nonlinear system with high-dimensional lifted states using data-driven methods~\cite{korda2018linear}. The Koopman-based EMPC framework solves optimization problems subject to system constraints defined by the established linear Koopman models. To form an economic objective function, the economic cost is approximated as a function of lifted states. In~\cite{han2024efficient}, a quadratic function was proposed to facilitate efficient online implementation. In~\cite{albalawi2023koopman}, economic costs were estimated using a linear function of lifted states.  While Koopman-based EMPC provides a solution for handling unavailable accurate first-principles models, one remaining crucial issue is the need for full-state measurements. However, in real applications, obtaining online measurements of all key variables through hardware sensors may be infeasible~\cite{yin2019subsystem}. 
% State estimation is required additionally to facilitate the online implementation of Koopman-based EMPC. 

The data-enabled predictive control (DeePC) framework offers a promising approach to bypass the need of system modeling \cite{coulson2019data}. According to Willems’ fundamental lemma~(\cite{willems2005note}), linear time-invariant systems can be represented non-parametrically using collected input and output trajectories.
% Any trajectory of a specific system can be characterized by the Hankel matrix constructed from finite-length input and output trajectories obtained under a persistently exciting input sequence. 
In~\cite{coulson2019data}, data-enabled predictive control (DeePC) was proposed based on Willems’ fundamental lemma. 
% DeePC adopts a receding horizon optimization strategy to generate optimal control actions. 
In~\cite{yang2015data, zhang2023dimension}, order reduction was performed on the Hankel matrix using singular value decomposition (SVD) to enhance its online computational efficiency. In~\cite{shang2024willems,zhang2025deep}, Willems' fundamental lemma was leveraged to address nonlinear systems. In~\cite{xie2023linear}, a data-driven economic MPC framework for linear systems was proposed to minimize economic cost dynamically. 
% However, this approach is limited to linear systems. 
However, incorporating general economic cost functions can lead to non-convex optimization problems, which will reduce the efficiency of online implementation.

In this work, we propose an economic data-enabled predictive control (economic DeePC) framework for nonlinear systems. Inspired by the economic cost function approximation design in \cite{han2024efficient}, the proposed economic DeePC is formulated as an input-output control scheme that handles the nonlinear economic cost function in a convex fashion without requiring a first-principles process model and full-state measurements. In particular, the nonlinear economic cost is approximated by a quadratic function of the transformed output. 
Key output variables are reconstructed from the transformed output to meet system constraints during online implementation.
The appropriate nonlinear mapping that makes the transformed outputs compliant with the Willems' fundamental lemma is represented by a trained neural network.

\section{Preliminaries and problem formulation}
\subsection{Notation}
% $\mathbb{R}$ denotes the set of real numbers. 
$\mathbb{N}_{>0}$ denotes the set of positive integers. $\mathbb{E}$ denotes the expectation. $\{x\}_{j}^{l}:=\big[x_j^\top, \dots, x_l^\top \big]^\top$ denotes a sequence that contains vector $x$ from the time instant $j$ to $l$. $x_{j|k}$ is the state vector for sampling instant $j$ obtained at time instant $k$. 
% $\circ$ represents the composition of functions. 
$||x||^2$ represents the square of the Euclidean norm of vector $x$. $A^+$ represents the pseudo-inverse of matrix $A$. 
$\text{diag}\left( \cdot \right)$ represents a diagonal matrix. $\text{exp}(\cdot)$ denotes the element-wise exponential operator.
% $\textbf{I}_{n}$ is an identity matrix which has dimension $n\times n$. $\bf{0}$ denotes a null matrix of appropriate dimensions. 
The superscript `d' denotes that the corresponding data is collected offline.

\subsection{Non-parametric system representation}
Consider a discrete linear time-invariant (LTI) system:
\begin{subequations}\label{deepc:linearmodel}
\begin{align}
    x_{k+1} &= Ax_k+Bu_k \label{deepc:linearmodel:1}\\
    y_k &= Cx_k+Du_k \label{deepc:linearmodel:2}
\end{align}
\end{subequations}
where $k$ denotes the sampling instant; $x\in\mathbb{X}\subseteq\mathbb{R}^{n_x}$ denotes the system state vector; $u\in\mathbb{U}\subseteq\mathbb{R}^{n_u}$ is the control input; $y\in\mathbb{Y}\subseteq\mathbb{R}^{n_y}$ is the system output; $k$ denotes the sampling instant; $A\in \mathbb{R}^{n_x\times n_x}$, $B\in \mathbb{R}^{n_x\times n_u}$, $C\in \mathbb{R}^{n_y\times n_x}$, and $D\in \mathbb{R}^{n_y\times n_u}$ are system matrices.

Let $\mathbf{u}_{T}^d:= \{u^d \}_1^{T}$ and $\mathbf{y}_{T}^d:= \{ y^d\}_{1}^{T}$ denote the $T$-step offline collected input and output sequences of system (\ref{deepc:linearmodel}), respectively. 
Next, we introduce the concept of persistent excitation and Willems' fundamental lemma.

\begin{definition}\label{deepc:def:pe} (Persistent excitation \cite{willems2005note}).
% [Persistent excitation (\cite{willems2005note})]\label{deepc:def:pe}
Let $T, L \in \mathbb{N}_{>0}$ and $T\geq L$. A Hankel matrix of depth $L$ is constructed based on the input sequence $\mathbf{u}_T^d$, defined as follows:
\begin{equation}\label{hankel}
    \mathscr{H}_L(\mathbf{u}_T^d)=\left[\begin{array}{c c c c }
                    u_1^d & u_2^d & \dots & u^d_{T-L+1} \\
                    u_2^d & u^d_3 & \dots & u^d_{T-L+2} \\
                    \vdots & \vdots &\ddots &\vdots \\
                    u^d_{L} & u^d_{L+1} & \dots & u^d_{T}
                    \end{array}\right]
\end{equation}
The sequence $\mathbf{u}_T^d$ is persistently exciting of order $L$, if $\mathscr{H}_L(\mathbf{u}_T^d)$ has full row rank.
\end{definition}

\begin{lemma} (Willems' fundamental lemma \cite{willems2005note}).
% [\makebox[0.8\linewidth]{Willems' fundamental lemma \cite{willems2005note}}]
Consider a controllable LTI system (\ref{deepc:linearmodel}). $\mathbf{u}_{T}^d$ is persistently exciting of order $L+n_x$. According to~\cite{willems2005note}, any $L$-step trajectories $\mathbf{u}_{L}:=\{u\}_{1}^{L}\in \mathbb R^{n_uL}$ and $\mathbf{y}_{L}:=\{y\}_1^{L}\in \mathbb R^{n_yL}$ are the input and output trajectories of system (\ref{deepc:linearmodel}) if
\begin{equation}\label{fl}
    \left[\begin{array}{c}
           \mathscr{H}_L(\mathbf{u}_{T}^d) \\
           \mathscr{H}_L(\mathbf{y}_{T}^d)
          \end{array}
    \right] g = 
    \left[\begin{array}{c}
         \mathbf{u}_{L} \\
         \mathbf{y}_{L} 
    \end{array}
    \right]
\end{equation}
holds for a column vector $g\in\mathbb{R}^{T-L+1}$.
\label{deepc:lem}\leavevmode\unskip
\end{lemma}

Lemma~\ref{deepc:lem} (\cite{willems2005note}) provides a non-parametric representation of~(\ref{deepc:linearmodel}) using historical input and output trajectories $\mathbf{u}_{T}^d$ and $\mathbf{y}_{T}^d$. 
% It shows that if $\mathbf{u}^d_T$ is persistently exciting of order $L+n_x$, then according to Definition~\ref{deepc:def:pe}, any input and output trajectories of system~(\ref{deepc:linearmodel}) with length $L$, denoted by $\mathbf{u}_{L}$ and $\mathbf{y}_{L}$, can be described by the linear combination of the Hankel matrices, $\mathscr{H}_L(\mathbf{u}_{T}^d)$ and $\mathscr{H}_L(\mathbf{y}_{T}^d)$, with a column vector $g$.

\subsection{Data-enabled predictive control}
\label{sec:deepc}
% Data-enabled predictive control (DeePC), originally developed based on Willems’ fundamental lemma, is an alternative receding horizon control framework~(\cite{coulson2019data}). 
% Data-enabled predictive control (DeePC) is a receding horizon control framework developed based on Willems’ fundamental lemma~(\cite{coulson2019data}). 
% It does not require explicit modeling of the underlying system. Instead, DeePC creates a non-parametric representation of the considered system based only on historical data.  
Data-enabled predictive control (DeePC) is a receding horizon control framework that does not require modeling of the underlying system~(\cite{coulson2019data}). Instead, it creates a non-parametric representation of the considered system based only on historical data.

In DeePC, future multi-step ahead predictions are described using partitioned Hankel matrices. Let $T_{ini}, N_p \in \mathbb{N}_{>0}$ and $L=T_{ini}+N_p$, the Hankel matrices $\mathscr{H}_L(\mathbf{u}^d_T)$ and $\mathscr{H}_L(\mathbf{y}^d_T)$ are divided into two segments: the past data and the future data, as follows \cite{coulson2019data}:
\begin{equation}\label{deepc:4}
    \left[\begin{array}{c}
            U_p \\
            U_f
          \end{array}
    \right] := \mathscr{H}_L(\mathbf{u}^d_T),\ \left[\begin{array}{c}
            Y_p \\
            Y_f
          \end{array}
    \right] := \mathscr{H}_L(\mathbf{y}^d_T)
\end{equation}

In~(\ref{deepc:4}), $U_p$ and $Y_p$ contain the past input and output data, which correspond to the first $n_u \times T_{ini}$ rows of $\mathscr{H}_L(\mathbf{u}^d_T)$ and the first $n_y \times T_{ini}$ rows of $\mathscr{H}_L(\mathbf{y}^d_T)$, respectively. Similarly, $U_f$ and $Y_f$ represent the future input and output data corresponding to the last $n_u\times N_p$ rows of $\mathscr{H}_L(\mathbf{u}^d_T)$ and the last $ n_y\times N_p$ rows of $\mathscr{H}_L(\mathbf{y}^d_T)$, respectively. 

At each time instant $k$, $T_{ini}$-step input and output sequences $\mathbf{u}_{ini, k}:=\{u\}_{k-T_{ini}}^{k-1}$ and $\mathbf{y}_{ini, k}:=\{y\}_{k-T_{ini}}^{k-1}$ are used to initialize the DeePC algorithm. Based on Lemma~\ref{deepc:lem}, $N_p$-step future input and output prediction sequences $\hat{\mathbf{u}}_k:=\{\hat{u}\}_{k|k}^{k+N_p-1|k}$ and $\hat{\mathbf{y}}_k:=  \{\hat{y}\}_{k|k}^{k+N_p-1|k}$ satisfy~(\ref{fl}). The DeePC optimization problem at time instant $k$ can be formulated as follows~(\cite{coulson2019data}):
\begin{subequations}\label{deepc:deepc_opt}
\begin{align}
        \min_{g_k, \hat{\mathbf{u}}_k, \hat{\mathbf{y}}_k} \  &\Vert \hat{\mathbf{y}}_k - \mathbf{y}^r_k \Vert_Q^2 + \Vert \hat{\mathbf{u}}_k -\mathbf{u}^r_k\Vert_R^2 \label{deepc:deepc_opt:prob}\\
        \text{s.t.} \quad &\left[\begin{array}{c}
            U_p \\
            Y_P \\
            U_f \\
            Y_f
          \end{array}
    \right] g_k = 
    \left[\begin{array}{c}
         \mathbf{u}_{ini,k} \\
         \mathbf{y}_{ini,k} \\
         \hat{\mathbf{u}}_k \\
         \hat{\mathbf{y}}_k
    \end{array} 
    \right]\label{deepc:deepc_opt:1} \\
    &\hat{u}_{j|k} \in \mathbb{U}, \quad j = k, \ldots, k+N_p-1\\
    &\hat{y}_{j|k} \in \mathbb{Y}, \quad j = k, \ldots, k+N_p-1 
\end{align}
\end{subequations}
 
In~(\ref{deepc:deepc_opt}), $\mathbf{u}^r_k:=\{u^r\}_k^{k+N_p-1}$ and $\mathbf{y}^r_k:=\{y^r\}_k^{k+N_p-1}$ are the input and output references, respectively; $Q\in \mathbb{R}^{n_yN_p\times n_yN_p}$ and $R\in \mathbb{R}^{n_uN_p\times n_uN_p}$ are the weighting matrices. $\hat{\mathbf{u}}_k$ and $\hat{\mathbf{y}}_k$ can be uniquely determined by $g_k$ through $\hat{\mathbf{u}}_k = U_f g_k$ and $\hat{\mathbf{y}}_k = Y_f g_k$ in (\ref{deepc:deepc_opt:1}). The first element $\hat u^*_{k|k}$ of the optimal input sequence $\hat{\mathbf{u}}_k^* = \big[\hat{u}^{* \top}_{k|k}, \ldots, \hat{u}_{k+N_p-1|k}^{* \top}\big]^\top$ is used as the system input which is to be applied to system~(\ref{deepc:linearmodel}) at time instant $k$ (see, e.g., \cite{coulson2019data}).

\subsection{Motivation and problem formulation}

In this work, we consider general discrete-time nonlinear systems in the following form:
\begin{subequations}\label{deepc:nlmodel}
    \begin{align}
        x_{k+1} &= f(x_k, u_k)\label{deepc:nlmodel:1}\\
        y_k &= h(x_k)\label{deepc:nlmodel:2}
    \end{align}
\end{subequations}
where
$f: \mathbb{X}\times\mathbb{U}\to\mathbb{X}$ is a nonlinear function that describes the dynamical behaviors of the system; $h: \mathbb{X}\to \mathbb{Y}$ is the output measurement function.

The dependence of the real-time economic operational cost on the system input and output is characterized by a nonlinear economic cost function $c_k$ as follows:
\begin{equation}\label{deepc:ecost}
    c_k = \ell_e(u_k,y_k) 
\end{equation}
where $\ell_e: \mathbb{Y}\times\mathbb{U}\to\mathbb{R}$.

In this work, we aim to leverage the DeePC framework to propose an economic data-enabled predictive control approach, which can be applied to minimize the operational cost while ensuring the satisfaction of hard constraints on system output. As is analogous to the relationship between set-point tracking MPC and EMPC, a potential solution to extend DeePC to economic DeePC is to replace the quadratic stage cost in~(\ref{deepc:deepc_opt:prob}) with economic cost function $\ell_e$; this modification has been implemented in a data-driven economic predictive control design for LTI systems in~\cite{xie2023linear}. However, since the economic cost function $\ell_e$ in (\ref{deepc:ecost}) is typically nonlinear, it compromises the convexity of the original DeePC formulation. In that case, solving the resulting non-convex optimization problem can be more computationally expensive than solving the convex optimization problem associated with the original DeePC. 
% Moreover, the computational complexity of the DeePC problem is influenced by the amount of historical data used to construct the Hankel matrix. As the amount of historical data increases, combined with the challenges posed by the nonlinear economic cost function $\ell_e$ in~(\ref{deepc:ecost}), the real-time implementation of straightforward economic DeePC formulations can become significantly constrained.

Building on the above considerations, we aim to develop an economic DeePC framework for the nonlinear system in (\ref{deepc:nlmodel}) that inherits the convexity of the online optimization problem in the original DeePC.

\section{Economic data-enabled predictive control approach}

In this section, we leverage machine learning to propose a convex economic data-enabled predictive control approach, referred to as economic DeePC, designed for economic control of nonlinear systems in~(\ref{deepc:nlmodel}).

% \subsection{Key components of the proposed training approach}
 \subsection{Training of the proposed economic DeePC}

% \begin{figure*}[t!]
%     \centering
%     \includegraphics[width=0.7\textwidth]{image/train.pdf}
%     \caption{An overview of the proposed economic DeePC training approach.}
%     \label{fig:train}
% \end{figure*}

\begin{figure}[t!]
    \centering
    \includegraphics[width=0.49\textwidth]{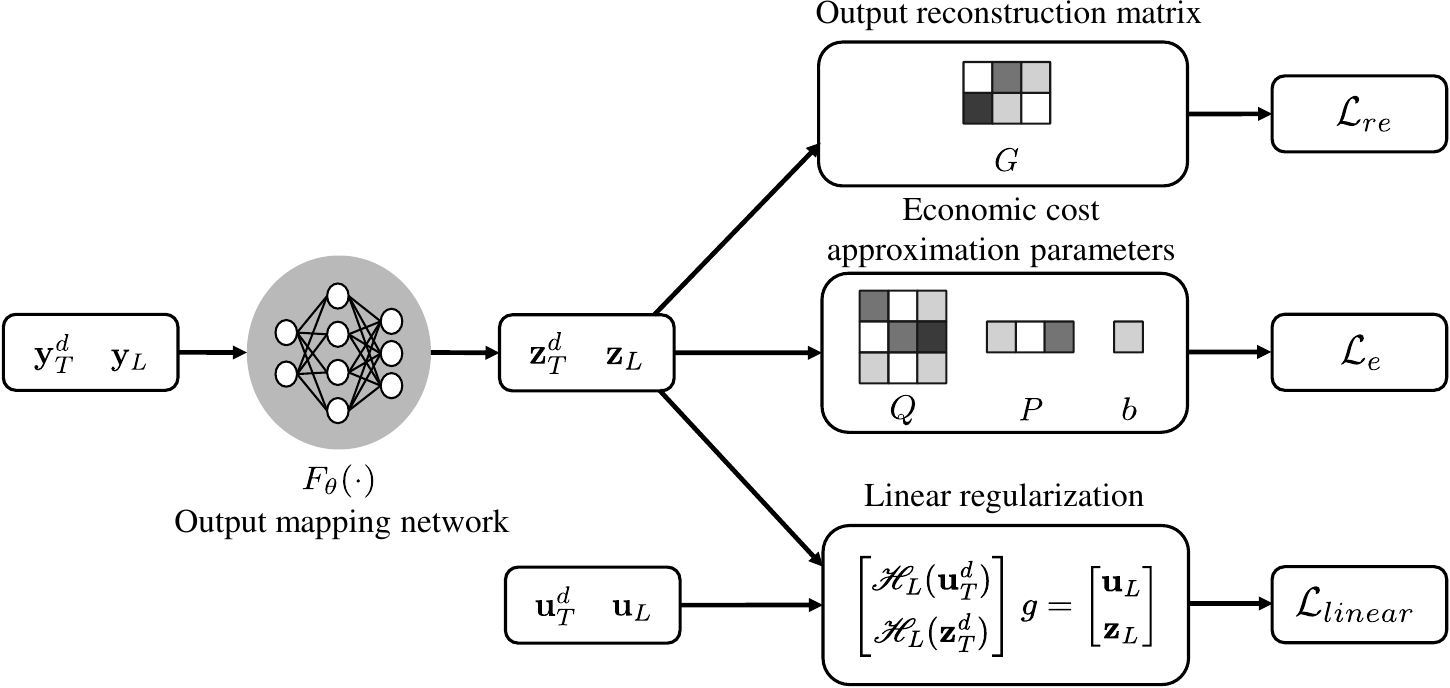}
    \caption{An illustration of the training of an economic DeePC controller based on the proposed method.}
    \label{fig:train}
\end{figure}

Firstly, we introduce the three key components of the proposed economic DeePC training process and formulate the optimization problem for the training process.

Fig.~\ref{fig:train} provides an illustration of the training process for the proposed machine learning-based economic DeePC approach. We elaborate on the three components that contain trainable parameters.
\begin{enumerate}
\item \emph{Neural network for output mapping}:
    A neural network, $F_\theta(\cdot):\mathbb{R}^{n_y} \to \mathbb{R}^{n_z}$, is introduced to map the original system output $y $ to vector $z\in \mathbb{R}^{n_z}$ in a new space. The parameters of $F_\theta(\cdot)$, denoted by $\theta$, are trainable.
    Two open-loop output trajectories of the system, $\mathbf{y}_T^d$ and $\mathbf{y}_L$, are reshaped into matrices 
    % \textcolor{blue}{$\mathbf{Y}_T^d = [y^d_1,\dots,y^d_T]^\top$ and $\mathbf{Y}_L = [y_1,\dots,y_L]^\top$}
    $\mathbf{Y}_T^d = [y^d_1,\dots,y^d_T]^\top \in \mathbb{R}^{T\times n_y}$ and $\mathbf{Y}_L = [y_1,\dots,y_L]^\top\in \mathbb{R}^{L\times n_y}$
    which are then fed into neural network $F_\theta(\cdot)$ 
    to generate their corresponding transformed output sequences, as follows:
    \begin{equation}\label{deepc:nn}
        \mathbf{Z}^d_{T} = F_\theta(\mathbf{Y}^d_T),\ \mathbf{Z}_L = F_\theta(\mathbf{Y}_L)
    \end{equation}
    where $\mathbf{Z}^d_{T} \in \mathbb{R}^{T\times n_z}$ and $\mathbf{Z}_{L} \in \mathbb{R}^{L\times n_z}$ are matrices of the transformed outputs. To facilitate the implementation of economic DeePC, we reshape transformed output matrices back into multi-step transformed output trajectories $\mathbf{z}^d_{T} = \{z^d\}_{1}^{T}$ and $\mathbf{z}_L = \{z\}_{1}^{L}$. 

\item \emph{Approximation of economic cost in quadratic form}: 

    The economic cost approximation follows the approximation method in the Koopman-based convex EMPC design in \cite{han2024efficient}, where a quadratic function was learned to approximate a non-convex economic state cost for a wastewater treatment process. Specifically, leveraging the approximation method from \cite{han2024efficient}, the transformed vector $z$ in the new space is used to create a quadratic expression to approximate the nonlinear economic cost function $\ell_e$ in (\ref{deepc:ecost}). An approximation of the economic cost for sampling instant $k$, denoted by $\hat c_k$, is computed as follows:
    \begin{equation}\label{deepc:c}
        \hat{c}_k = z_k^\top Qz_k + Pz_k +b
    \end{equation}
    where $Q\in\mathbb{R}^{n_z\times n_z}$ is a positive-definite matrix; $P\in\mathbb{R}^{1\times n_z}$ is a real-valued vector; $b\in\mathbb{R}$ is a real-valued scalar.
    To ensure the positive definiteness of $Q$, we require that $Q$ takes the form of 
    $Q = \text{diag}\big(\text{exp}(q)\big)$, where $q\in\mathbb{R}^{n_z}$ is a real-valued vector. $q$, $P$, and $b$ are trainable parameters.
\item \textbf{Output reconstruction matrix}: 
    The vector of output variables on which hard constraints need to be imposed on is denoted by $y_k^c$. A trainable output reconstruction matrix $G\in\mathbb{R}^{n_c\times n_z}$ is used to reconstruct $y_k^c$ from the transformed output $z_k$ to ensure that the system constraints are satisfied during system operation.
    % , where $\mathbb{Y}_c \subseteq \mathbb{Y}$ is the constrained system output space. 
    The reconstructed system output can be calculated as follows:
    \begin{equation}\label{deepc:re}
        \hat{y}_k^c = Gz_k
    \end{equation}
    where $\hat{y}_k^c$ is an approximation of the output vector $y^c_k$ at sampling instant $k$. 
\end{enumerate}

Next, we present the optimization problem associated with the training process for the proposed economic DeePC method. Given an open-loop data set $\mathcal{D}:=\{\mathbf{u}^d_T,\mathbf{u}_L,\mathbf{y}^d_T,\mathbf{y}_L, \mathbf{c}^d_T,\mathbf{c}_L\}$, where $\mathbf{c}_T^d:=\{c^d\}_1^{T}$ and $ \mathbf{c}_L:=\{c\}_1^{L}$ are sequences of actual economic costs computed using the corresponding system output data in $\mathbf{y}^d_T$ and $\mathbf{y}_L$ and control input data contained $\mathbf{u}^d_T$ and $\mathbf{u}_L$ based on~(\ref{deepc:ecost}), the objective is to optimize the trainable parameters ($\theta$, $q$, $P$, $b$, and $G$). 
% \textcolor{blue}{All trajectories used for training are open-loop data.}
Consequently, the optimization problem for the training of the proposed economic DeePC method is formulated as follows:
\begin{equation}\label{deepc:op_train}
        \min_{\theta, q,P,b,G}\ \mathcal{L} = \min_{\theta, q,P,b,G}\  \alpha_1 \mathcal{L}_{e} + \alpha_2 \mathcal{L}_{re}  + \alpha_3  \mathcal{L}_{linear}
\end{equation}
where $\mathcal{L}_{e}$ denotes the loss for approximating the economic cost; $\mathcal{L}_{re}$ represents the loss for reconstructing the output variables that required to satisfy hard constraints; $\mathcal{L}_{linear}$ represents the loss that quantifies the extent of violation of Willems' fundamental lemma; $\alpha_i,i=1,2,3$, are weights for three different losses.

Specifically, on the right-hand side of~(\ref{deepc:op_train}), $\mathcal{L}_{e}$ computes the discrepancy between the actual economic cost and the estimated cost, given as follows:
\begin{equation}
\label{deepc:lcost}
    \mathcal{L}_{e} = \mathbb{E}_{\mathcal{D}}\ \left(||\mathbf{c}_T^d - \hat{\mathbf{c}}_T^d||^2 + ||\mathbf{c}_L - \hat{\mathbf{c}}_L||^2\right)
\end{equation}
where $ \hat{\mathbf{c}}_T^d:=\{\hat c^d\}_1^{T}$ and $ \hat{\mathbf{c}}_L:=\{\hat c\}_1^{L}$ denote the sequences of the approximated economic cost obtained based on the transformed output $\mathbf{z}^d_T$ and $\mathbf{z}_L$ following~(\ref{deepc:c}), respectively. 

$\mathcal{L}_{re}$ in the second term on the right-hand side of~(\ref{deepc:op_train}) accounts for the discrepancy between the ground-truth and the reconstructed system outputs on which hard constraints are imposed; it is with the following form:
\begin{equation}
\label{deepc:lstate}
\mathcal{L}_{re} = \mathbb{E}_{\mathcal{D}}\  \left(\big|\big| \mathbf{y}^{c,d}_T - \hat{\mathbf{y}}^{c,d}_T \big|\big|^2+ \big|\big|  \mathbf{y}^{c}_L - \hat{\mathbf{y}}^{c}_L   \big|\big|^2\right)
\end{equation}
where $\mathbf{y}^{c,d}_T:=\{y^{c,d}\}_0^{T-1}$ and $\mathbf{y}^{c}_L:=\{y^{c}\}_0^{L-1}$ contain the elements of $\mathbf{y}_T^d$ and $\mathbf{y}_L$ on which hard constraints should be imposed; $\hat{\mathbf{y}}^{c,d}_T:=\{\hat y^{c,d}\}_0^{T-1}$ and $\hat{\mathbf{y}}^{c}_L:=\{\hat y^{c}\}_0^{L-1}$ represent the reconstructed output calculated based on~(\ref{deepc:re}). 

$\mathcal{L}_{linear}$ in the third term on the right-hand side of~(\ref{deepc:op_train}) is in the following form:
\begin{subequations}
\label{deepc:llinear}
\begin{align}
    \mathcal{L}_{linear} = \ &\mathbb{E}_{\mathcal{D}}\ ||\mathbf{z}_L- \mathscr{H}_L(\mathbf{z}_T^d)g||^2 \\
    = \ &\mathbb{E}_{\mathcal{D}}\ ||\mathbf{z}_L- \mathscr{H}_L(\mathbf{z}_T^d)\mathscr{H}_{L}(\mathbf{u}_T^d)^+\mathbf{u}_L||^2
\end{align}
\end{subequations}
This term is to guide the training of $F_\theta(\cdot)$ in a way that the transformed outputs, which are generated by $F_\theta(\cdot)$, conform to the Willems' fundamental lemma.

% After the training process using open-loop data, the optimal parameters, denoted by $\theta^*$, $Q^*$, $P^*$, $b^*$, and $G^*$, will be used for online economic data-enabled predictive control design and implementation. 

% \begin{remark}
% In the proposed economic DeePC approach, a neural network $F_\theta(\cdot)$ is utilized to map the original system output to a transformed output. The transformed output serves three key purposes: 1) it facilitates the establishment of a quadratic approximation of the nonlinear economic cost function using trainable parameters; 2) it aids in reconstructing key output variables on which hard constraints need to be imposed; 3) it facilitates the compliance with Willems' fundamental lemma.
% \end{remark}

% \begin{remark}
% From a practical perspective, the variables that need to satisfy hard constraints need to be measured online, and therefore can be treated as part of the system output.
% \end{remark}

\subsection{Economic data-enabled predictive control}

The economic cost approximation follows the approximation method in the Koopman-based convex EMPC design in \cite{han2024efficient}, where a quadratic function was learned to approximate a non-convex economic state cost for a wastewater treatment process. Specifically, leveraging the economic cost approximation design from \cite{han2024efficient}, the transformed vector $z$ in the new space is used to create a quadratic expression to approximate the nonlinear economic cost function $\ell_e$ in (\ref{deepc:ecost}).

Once the training is completed, the optimal parameters $\theta^*$, $Q^*$, $P^*$, $b^*$, and $G^*$ are used to formulate the economic DeePC design. 
Inspired by~\cite{coulson2019data}, the online optimization problem for the proposed economic DeePC is formulated as follows:
\begin{subequations}\label{deepc:keconomic DeePC_opt}
\begin{align}
    \min_{g_k} \sum_{j=k}^{k+N_p-1}\ & \beta(z_{j|k}^\top Q^*z_{j|k} + P^*z_{j|k}+b^*) + (\Delta \hat{u}_{j|k}^\top R\Delta \hat{u}_{j|k})\label{deepc:keconomic DeePC_opt:prob}\\
    \text{s.t.}\ &\left[\begin{array}{c}
            U_p \\
            Z_p \\
            U_f \\
            Z_f
          \end{array}
    \right]\ g_k = 
    \left[\begin{array}{c}
         \mathbf u_{ini,k} \\
         \mathbf z_{ini,k} \\
         \hat{\mathbf{u}}_k \\
         \hat{\mathbf{z}}_k
    \end{array}
    \right]\label{deepc:keconomic DeePC_opt:1}\\
    % \hat{x}^c_{j|k} &= G^*\hat{z}_{j|k}\label{deepc:keconomic DeePC_opt:2}\\
    &\hat{u}_{j|k}\in\mathbb{U},\ j = k,\dots,k+N_p-1\label{deepc:keconomic DeePC_opt:2} \\
    % \hat{x}^c_{j|k} &\in \mathbb{X}_c,\ j = k,\dots,k+N_p-1\label{deepc:keconomic DeePC_opt:3}
    &G^*\hat{z}_{j|k} \in \mathbb{Y}_c,\ j = k,\dots,k+N_p-1\label{deepc:keconomic DeePC_opt:3}
\end{align}
\end{subequations}
where $N_p$ represents the prediction horizon; $R\in \mathbb{R}^{n_u\times n_u}$ is a positive-definite matrix; $\beta\in \mathbb{R}$ is a positive scalar;
$\hat{\mathbf{z}}_k:=  \{\hat{z}\}_{k|k}^{k+N_p-1|k}$ denotes the predicted transformed output sequence; 
$\mathbb{Y}_c$ is the output space of the output vector $\hat{y}^c_k$ on which hard constraints need to be imposed;
$\Delta \hat{u}_{j|k} = \hat{u}_{j|k}-\hat{u}_{j-1|k}$ is the rate of change in the control input;
$Z_p$ and $Z_f$ are the partitions of the Hankel matrix constructed by the transformed offline collected output, that is, $[Z_p^\top, Z_f^\top]^\top := \mathscr{H}_{T_{ini}+N_p}(\mathbf{z}_T^d)$; 
$\mathbf z_{ini,k}$ and $\mathbf{z}^d_T$ are the trajectories of the initial transformed output and the offline transformed output generated using $\mathbf y_{ini,k}$ and $\mathbf{y}^d_T$ according to~(\ref{deepc:nn}).

% \begin{figure}[t]
%     \centering
%     \includegraphics[width=0.49\textwidth]{image/edeepc.pdf}
%     \caption{The proposed economic DeePC design.}
%     % \textcolor{blue}{Control policy to control action.}}
%     \label{fig:keconomic DeePC}
% \end{figure}

% In~(\ref{deepc:keconomic DeePC_opt}), (\ref{deepc:keconomic DeePC_opt:prob}) is the objective function considers economic costs and the penalty on the rate of change in the system input; (\ref{deepc:keconomic DeePC_opt:1}) is the non-parametric representation based on the transformed output $z$ and control input $u$; (\ref{deepc:keconomic DeePC_opt:2}) and (\ref{deepc:keconomic DeePC_opt:3}) are the imposed hard constraints of the system input and the reconstructed output.

The optimal control input sequence $\hat{\mathbf{u}}_k^*$ can be obtained based on the optimized DeePC operator $g_k^*$ as follows:
\begin{equation}\label{deepc:optu}
    \hat{\mathbf{u}}_k^* = U_fg_k^*
\end{equation}
where $\hat{\mathbf{u}}_k^* = \big[\hat{u}^{* \top}_{k|k}, \ldots, \hat{u}_{k+N_p-1|k}^{* \top} \big]^\top$. The first element $\hat{u}^*_{k|k}$ of this optimal control sequence is applied to system~(\ref{deepc:nlmodel}) for real-time control.

\subsection{Reduced-order economic DeePC}

The dimension of $g_k$ in (\ref{deepc:keconomic DeePC_opt}) is dependent on the size of the Hankel matrix, which is typically large. To reduce the computation time for solving the optimization problem (\ref{deepc:keconomic DeePC_opt}) during online implementation, we incorporate the singular value decomposition (SVD)-based dimension reduction method in~\cite{zhang2023dimension} into the proposed economic DeePC, to achieve faster online control implementation. 

A reduced-order matrix can be established to approximate the Hankel matrix and form a corresponding reduced-ordered DeePC operator $\bar{g}_k$. Based on SVD~(\cite{wall2003singular}), the Hankel matrix can be represented as follows:  
\begin{equation}\label{deepc:svd}
% \mathscr{H}_{T_{ini}+N_p} =
\left[\begin{array}{c}
     \mathscr{H}_{T_{ini}+N_p}(\mathbf u^d_T) \\
    \mathscr{H}_{T_{ini}+N_p}(\mathbf z^d_T)
    \end{array}
\right] = \left[W_1,W_2\right]\left[\begin{array}{cc}
    \Sigma_1&\bf{0} \\
    \bf{0}&\bf{0}
\end{array}
\right]\left[V_1,V_2\right]^{\top}
\end{equation}
% \textcolor{red}{Unify the Hankel matrix symbol in this section, use $\mathcal{H}_{L}$ or $\mathcal{H}_{T_{ini}+N_p}$ in (\ref{xxxxxc})?}
where $\Sigma_1\in\mathbb{R}^{n_r\times n_r}$ is the diagonal matrix of non-zero singular values;
% values that includes the top $h$ non-zero singular values; 
$W :=\left[W_1,W_2\right]$ and $V:=\left[V_1,V_2\right]$ are singular vector matrices satisfying $WW^\top = W^\top W = \textbf{I}_{(n_u+n_y)L}$ and $VV^\top = V^\top V = \textbf{I}_{T-T_{ini}-N_p+1}$. The reduced-order approximation $\bar{\mathscr{H}}_{T_{ini}+N_p}$ can be established as follows~(\cite{zhang2023dimension}):
\begin{equation}\label{svd:hankel}
\bar{\mathscr{H}}_{T_{ini}+N_p} = W_1\Sigma_1
\end{equation}
% \textcolor{blue}{where $\bar{\mathcal{H}}_L(\mathbf{u}^d_T) \in \mathbb{R}^{xx \times xx}$ is the reduced-order Hankel matrix. The dimension of g is reduced from xx to h.}
where $\bar{\mathscr{H}}_{T_{ini}+N_p} \in \mathbb{R}^{(n_u+n_y)(T_{ini}+N_p) \times n_r}$ is the reduced-order Hankel matrix. The number of columns of the Hankel matrix and the dimension of $\bar{g}_k$ is reduced from $T-T_{ini}-N_p+1$ to $n_r$.

The optimization problem of the proposed reduced-order economic DeePC can be formulated as follows:
\begin{subequations}\label{deepc:svd_opt}
\begin{align}
    \min_{\bar g_k} \sum_{j=k}^{k+N_p-1}\ & \beta(z_{j|k}^\top Q^*z_{j|k} + P^*z_{j|k}+b^*) + (\Delta \hat{u}_{j|k}^\top R\Delta \hat{u}_{j|k})\label{deepc:svd_opt:prob}\\
    \text{s.t.}\ &\bar{\mathscr{H}}_{T_{ini}+N_p} \bar g_k = 
    \left[\begin{array}{c}
         \mathbf u_{ini,k} \\
         \mathbf z_{ini,k} \\
         \hat{\mathbf{u}}_k \\
         \hat{\mathbf{z}}_k
    \end{array}
    \right]\label{deepc:svd_opt:1}\\
    % &\hat{x}^c_{j|k} = G^*\hat{z}_{j|k}\label{deepc:svd_opt:2}\\
    &\hat{u}_{j|k}\in\mathbb{U},\ j = k,\dots,k+N_p-1 \label{deepc:svd_opt:3}\\
    % &\hat{x}^c_{j|k} \in \mathbb{X}_c,\ j = k,\dots,k+N_p-1\label{deepc:svd_opt:4}
    &G^*\hat{z}_{j|k} \in \mathbb{Y}_c,\ j = k,\dots,k+N_p-1\label{deepc:svd_opt:4}
\end{align}
\end{subequations}
% In~(\ref{deepc:svd_opt}), the dimension of $\bar{g}_k$ is $n_r$. 

% The online implementation process of the reduced-order economic DeePC is illustrated in Algorithm \ref{alg:cap}.

\section{Simulation results}

\subsection{Process description and operation objective}
% A schematic of the simulated chemical process is given in Fig.~\ref{fig:cstr}. 
The studied process comprises two continuous-stirred tank reactors connected in series.
In this process, two irreversible second-order reactions that convert reactant $\text{A}$ to the desired product $\text{B}$ take place simultaneously. A schematic of the process and the detailed process descriptions are referred to~\cite{wu2020economic}.

The state variables include the concentration of material $\text A$ and the temperature in each of the reactors. The state vector $x = [C_{\text{A}1}, T_1, C_{\text{A}2}, T_2]^\top$, where $C_{\text{A}i}$ represents the concentration of material $\text A$ in the $i$th reactor, $T_i$ is the temperature in the $i$th reactor, $i=1,2$. The input vector includes the concentration of reactant $\text A$ in the feed inlets to the two reactors $C_{\text{A}10}$ and $C_{\text{A}20}$, and the heating input rates for the two reactors $Q_1$ and $Q_2$, that is, $u = [C_{\text{A}10}, Q_1, C_{\text{A}20}, Q_2]^\top$. For this process, we consider that the output vector is the same as the state vector.

The economic profit, as is adopted from~\cite{wu2020economic}, is as follows:
\begin{equation}
    \ell_e(u_k,y_k) = k_0e^{\frac{-E}{RT_1}}C^2_{\text{A}1} + k_0e^{\frac{-E}{RT_2}}C^2_{\text{A}2}
\end{equation}
% The control objective is to maximize the economic profit. From a practical perspective, establishing an accurate first-principles dynamic model for chemical reaction processes can be challenging.
% The model in~(\ref{edeepc: cstr:equation}) is only used as a simulator. 
% In this case study, we aim to apply the proposed method to build a data-driven economic predictive control scheme for this chemical process.

\subsection{Simulation settings}
\subsubsection{Control methods and parameters}

    \begin{table}[t]
    \renewcommand\arraystretch{1.25}
    \caption{Hyperparameters for training.} \label{table:trainingparameter}
    \centering %\small
    \begin{tabular}{ c c }
          \toprule
           Parameters& Values \\
          \midrule
          % Length of trajectory $\mathbf u^d_T,\mathbf y^d_T$: $T$ & $1\times10^3$/$1\times10^3$\\
          % Length of trajectory $\mathbf u_{L},\mathbf y_{L}$: $L$ & $1\times10^3$/$9\times10^3$ \\\
          % Transformed output dimension: $n_z$ & 10 \\
          Dimension of hidden states  & 128 \\
          Number of hidden layers & 2 \\
          Activate function & ReLU\\
          % Training data percentage & $70\%$ \\
          % Validation data percentage & $20\%$ \\
          % Test data percentage & $10\%$ \\
          Training epoch & 100\\
          Batch size & 100\\
          % Optimizer & Adam\\
          \bottomrule
    \end{tabular}
    \end{table}

We consider two data-driven economic predictive control methods: the economic DeePC proposed in this work, and the learning-based Koopman EMPC in~\cite{han2024efficient}.

The following parameters are used for the proposed method: $T = 10^3$, $T_{ini} = 2$, $N_p=5$, and $n_z=10$. The hyperparameters for training the neural network in the economic DeePC framework are provided in Table~\ref{table:trainingparameter}. Additionally, in this chemical process, we aim to maximize profit instead of minimize operating costs. Therefore, matrix $Q$ is set to be negative definite in the training phase to facilitate the formulation of a convex optimization problem, that is, $Q = \text{diag}(- \text{exp}(q))$. 
% Consequently, the online optimization problem minimizes the negative profit function, which results in a convex optimization problem.

For the learning-based Koopman EMPC proposed in~\cite{han2024efficient}, the Koopman model is trained over 100 epochs. The dimension of the lifted state is set to 10.
% with a learning rate of $10^{-5}$. The dimension of the lifted state is set to 10, and the prediction horizon is 30. 
% The neural network 
Other hyperparameters follow the configuration outlined in~\cite{han2024efficient}. The control objective function is the same as that of the economic DeePC.

\subsubsection{Data generation}

First, open-loop simulations are conducted to generate data for constructing the Hankel matrices and training the neural networks. The sampling period is 0.025 hours. The control inputs are bounded by $1.5~ \text{kmol}\cdot\text{m}^{-3} \leq C_{\text{A}i0} \leq  6.5~ \text{kmol}\cdot\text{m}^{-3}$ and $-10^4~ \text{kJ}\cdot\text{h}^{-1} \leq  Q_i \leq  10^5~ \text{kJ}\cdot\text{h}^{-1}$, $i=1,2$. Bounded random disturbances are added to the process. Specifically, stochastic disturbances added to mass fractions are generated following Gaussian distribution $\mathcal N(0,0.01)$ and then made bounded within a range of $[-1,1]$. Stochastic disturbances added to temperatures are generated following Gaussian distribution $\mathcal N(0,1)$ and then made bounded within a range of $[-50,50]$.  

In training, we consider two cases to evaluate the control methods: Case 1 with $2 \times 10^3$ data samples and Case 2 with $10^4$ samples. Datasets are divided into training, validation, and test data in a ratio of 7:2:1. In both cases, economic DeePC uses $10^3$ samples to construct the Hankel matrix and uses the remaining data for training. The learning-based Koopman EMPC utilizes all the samples to train the neural network. Trainable parameters in both method are optimized using Adam~(\cite{kingma2014adam}).
% The data used for training the neural networks of the two control methods are then divided into training, validation, and test data with the proportion of 70\%, 10\%, and 20\%, respectively. 

\subsection{Control performance}

\begin{figure}[!t]
    \centering
    \includegraphics[width=0.49\textwidth]{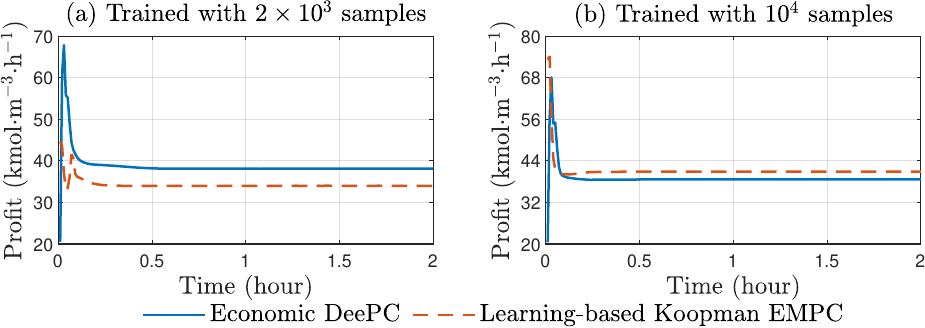}
    \caption{The economic profits for the closed-loop chemical process under the proposed economic DeePC and learning-based Koopman EMPC in \cite{han2024efficient}.}
    \label{fig:result}
\end{figure}

% \begin{table}[!t]
%   \renewcommand\arraystretch{1.25}
%   \caption{\textcolor{blue}{The results of average economic profits under two control schemes using optimal parameters trained with different sample sizes.}}\label{table:avgcost}\vspace{2mm}
%   \centering %\small
%     \begin{tabular}{ c c c c}
%       \toprule
%        Sample size & $2\times 10^3$ & $10^4$ \\
%       \midrule
%       Economic DeePC & $38.7502$ & $39.0457$ \\
%       Koopman-based EMPC & $34.2901$ & $41.2660$\\
%       \bottomrule
%     \end{tabular}
% \end{table}

% \begin{table}[!t]
%   \renewcommand\arraystretch{1.25}
%   \caption{The results of average economic profits under two control schemes using optimal parameters trained with different sample sizes.}\label{table:avgcost}\vspace{2mm}
%   \centering %\small
%     \begin{tabular}{|c|c|c|}
%     \hline
%     \multirow{2}{*}{\thead{Sample \\ size}} & \multicolumn{2}{c|}{\thead{Average profit (kmol/m$^3$h)}} \\
%     \cline{2-3}
%     & \thead{Economic DeePC} & \thead{Koopman-based EMPC} \\
%     \hline
%     $2 \times 10^3$ & 38.7502 & 34.2901 \\
%     \hline
%     $10^4$ & 39.0457 & 41.2660 \\
%     \hline
%     \end{tabular}
% \end{table}

\begin{table}[!t]
  \renewcommand\arraystretch{1.05}
  \caption{The results of average economic profit (with unit kmol$\cdot$m$^{-3}\cdot$h$^{-1}$).}\label{table:avgcost}
  % \vspace{2mm}
  \centering %\small
    \begin{tabular}{ccc}
    \toprule
    & $2\times 10^3$ samples & $10^4$ samples\\
    \midrule
    Economic DeePC & 38.7502 & 39.0457 \\
    {\thead{\normalsize Learning-based \\  \normalsize Koopman EMPC \\ \normalsize (\cite{han2024efficient})}} &\raisebox{-0.4ex}{34.2901} &\raisebox{-0.4ex}{41.2660} \\
    \bottomrule
    \end{tabular}
\end{table}

% Figure~\ref{fig:result} presents the average economic profits obtained from two control methods under datasets with two different sample sizes. 
Fig.~\ref{fig:result} presents the trajectories of the averaged economic profits obtained based on the two methods with different data sizes. The results are generated based on 20 repeated simulations, respectively. As shown in Fig.~\ref{fig:result}(a), when $2\times 10^3$ samples are used, the proposed control method provides better economic performance as compared to the learning-based Koopman EMPC in \cite{han2024efficient}. When $10^4$ samples are used, the two methods provide comparable performance, with the learning-based Koopman EMPC (\cite{han2024efficient}) providing a slightly higher profit, as shown in Fig.~\ref{fig:result}(b).
% To quantitatively assess the control performance, we calculate 
The average economic profits for both methods under two different data sizes are shown in Table~\ref{table:avgcost}. 
Based on the simulation results, the proposed economic DeePC method shows better data efficiency.

% The average profits for the proposed method and the learning-based Koopman EMPC are based on 20 repeated simulations, respectively.
% Both control methods achieve higher economic profits as the amount of training data increases. The proposed economic DeePC method demonstrates high data efficiency.

\section{Conclusion}
We proposed a convex economic data-enabled predictive control method. A neural network was utilized to map the original system output to a transformed output vector, which was used to build a quadratic approximation of the nonlinear economic cost function. The output variables were reconstructed from the transformed output using a trainable output reconstruction matrix. With the reconstructed outputs, hard constraints on the system output were explicitly addressed. The trainable parameters, including the neural network parameters, the coefficients in the quadratic approximation of the economic cost, and the constrained output reconstruction matrix, were learned using open-loop system data. 
A simulated chemical process was used to illustrate the proposed method.

\end{document}